\documentstyle[11pt,newpasp,twoside,epsf]{article}
\def\edcomment#1{\iffalse\marginpar{\raggedright\sl#1\/}\else\relax\fi}
\reversemarginpar

\begin{document}
\title{The Massive Stellar Population in a Nearby Massive Low 
Surface Brightness Galaxy}
 \author{Patricia M.\ Knezek}
\affil{Space Telescope Science Institute, 3700 San Martin Drive,
Baltimore, MD, 21218, USA}
%\author{Ima Co-Author}
%\affil{The Name of My Institution, The Full Address of My Institution}

\begin{abstract}
We have observed the closest known large, low surface
brightness (LSB) galaxy, UGC~2302, with HST using WFPC2 and the
F555W and F814W filters. UGC~2302 is a typical large
LSB galaxy, and thus represents a good choice for detailed
stellar population studies.   
Large LSBs represent an extreme star formation
environment, both locally within their disks, and on a larger
scale, often residing in underdense regions of the universe. 
Resolving the underlying stellar component of such a galaxy
can provide critical information on how stars form and evolve
in low metallicity, yet gas-rich environments.
We resolve for the first time the
massive stars and tip of the red giant branch in such a
system.  
\end{abstract}

\section{Introduction}

The mere existence of low surface brightness disk galaxies (LSBs) has profound 
implications for the evolution of the universe, both at high and low redshifts. 
The number of LSBs found as a function of magnitude is consistent with no 
turnover in the counts down to the limit of current data (O'Neil \& Bothun 2000, 
Ferguson \& McGaugh 
1995), and thus their contributions to the galaxy luminosity and mass functions 
on both the low and high luminosity and mass ends are not well constrained.  
Little is known about their epoch of formation. 

Gas-rich, large LSBs also provide a 
window into the extreme environment of very inefficient star formation.  
Enormous efforts have been made to study blue compact dwarf galaxies (BCDs) and 
starburst galaxies as examples of  
extreme environments for vigorous star formation, while efforts to study the
other extreme have centered largely on dwarf galaxies. Only recently have the 
larger, gas-rich systems come under scrutiny.
%Many 
%questions about these galaxies remain to be answered.  What is their epoch of 
%formation?  How does star formation proceed in such quiescent, but massive, disk 
%galaxies, where perhaps the star formation rate is driven by neither 
%galaxy-galaxy interactions nor by spiral density waves?  What is the role of 
%neutral gas, and of dust (if any)?  Is the initial mass function (IMF) the same?  
%Are the feedback mechanisms the same?  Does star formation propagate?  What are 
%the star formation histories of gas-rich LSBs?  How do they differ from their 
%dwarf and HSB counterparts?
Yet despite this recent research activity, the 
underlying stellar populations of disk LSBs and their detailed star formation 
histories remain poorly understood.  Star formation in large LSBs has 
presumably proceeded in a much 
different way than in dwarf galaxies or HSB disk galaxies.  While broadband 
optical and near-infrared observations have begun with the intention of 
studying the stellar populations in the galaxies and comparing them to current 
models of varying metallicities, disentangling age versus metallicity effects 
using broadband filters remains problematical, even for well-studied HSB 
galaxies (Worthey 1994; Bruzual \& Charlot 1993; Charlot et al. 1996).   By 
resolving the brightest stars and the tip of the Red Giant Branch (TRGB), both 
in the relatively red central region and out in the blue disk of a large LSB, 
UGC~2302 ($D \sim 15$ Mpc), 
we obtain direct information on the age and metallicity of the underlying 
stellar population as a function of radius.

%\section{Observations}
%Observations taken in F555W and F184W
%Exposure times 15000 sec in F555W and 15600 sec in F814W
%Observations spaced over three epochs to search for Cepheid candidates

\begin{figure}
\plottwo{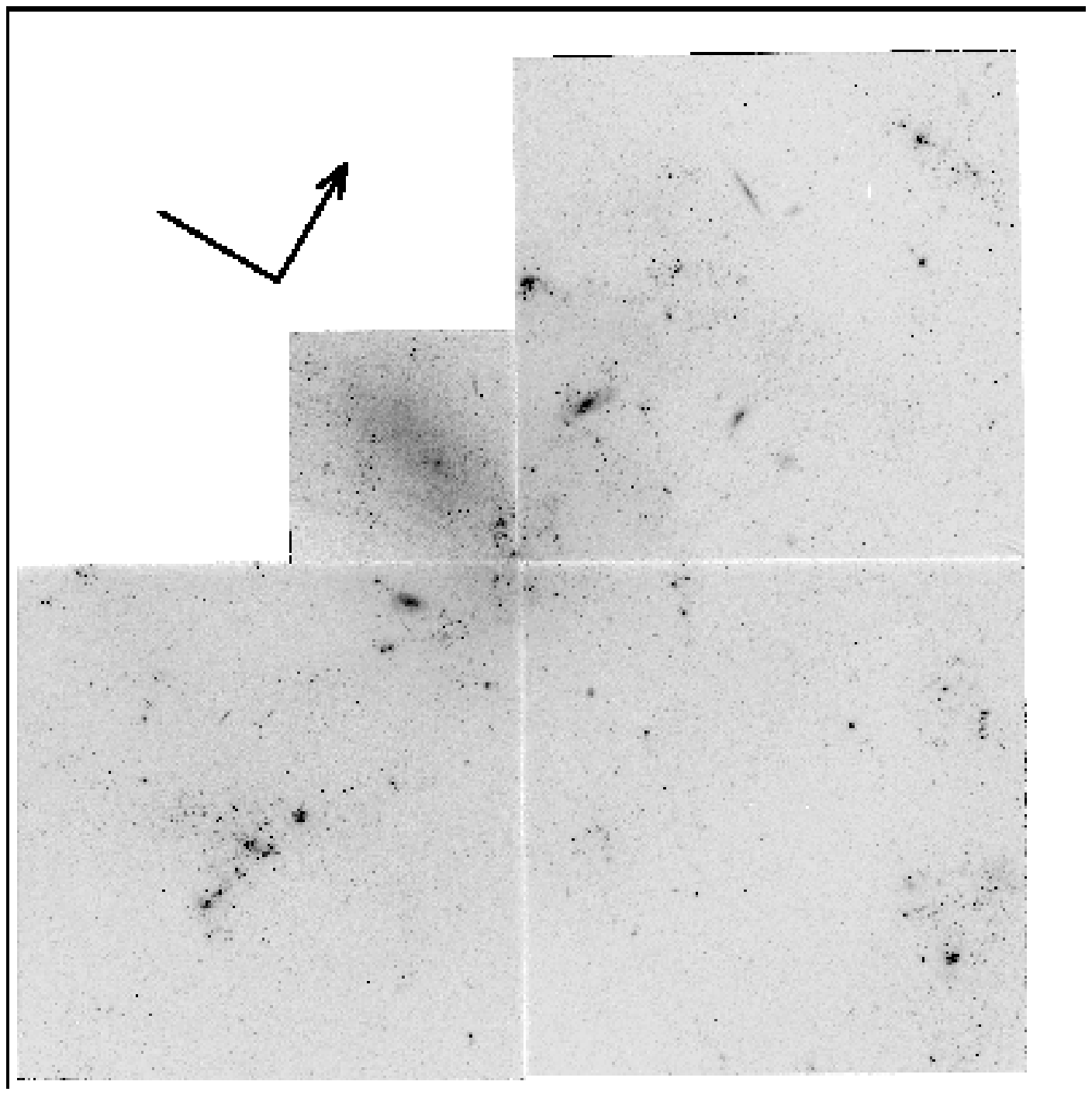}{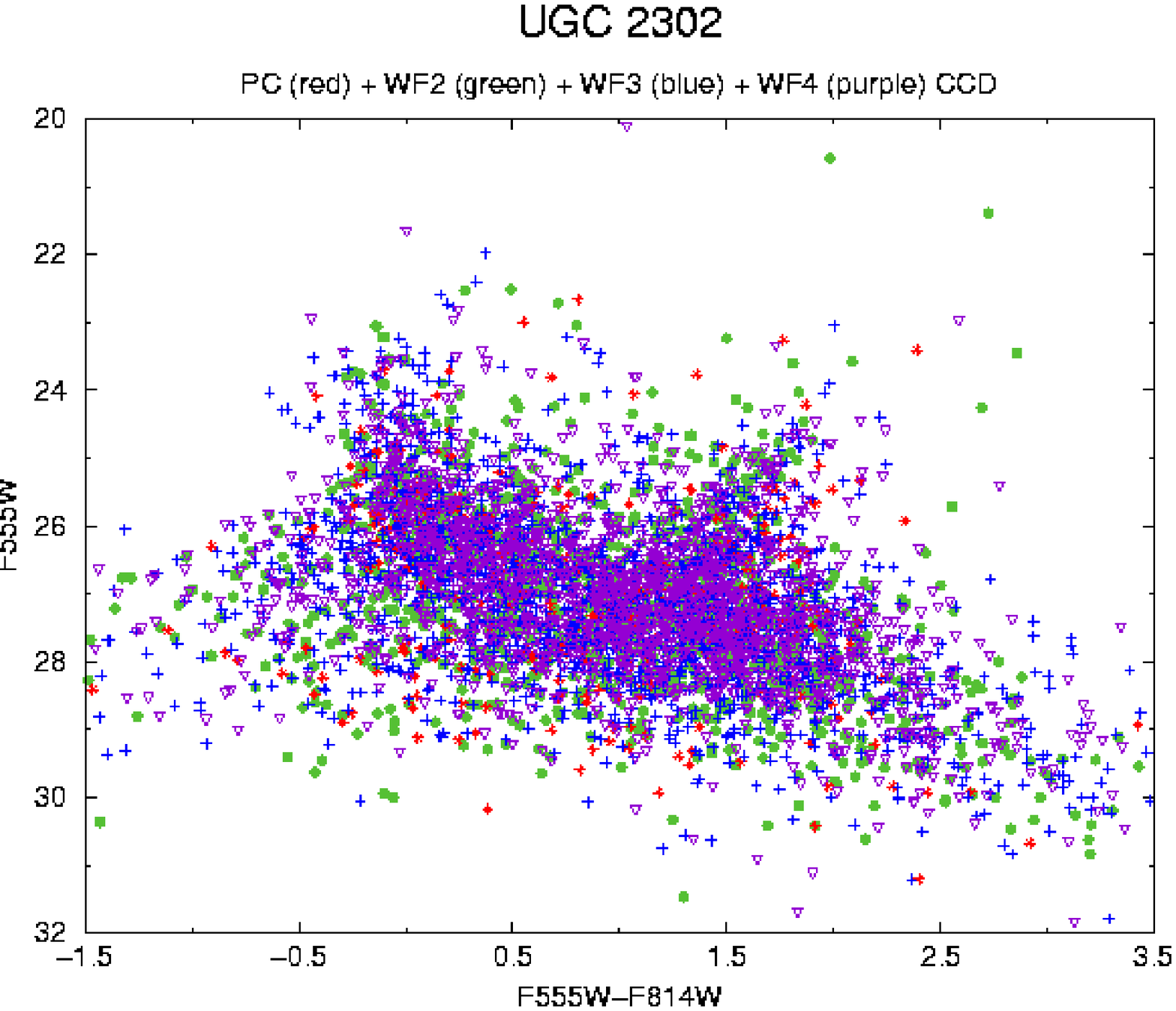}
\caption{The left figure is the F555W image of UGC~2302.  The arrow indicates 
North, while the solid line indicates East.  The right figure shows 
the color magnitude diagram, (m$_{F555W}$, m$_{F555W}-$m$_{F814W}$), for all four 
chips. The different colors indicate the different WFPC2 chips.}
\end{figure}

\section{Preliminary Results}

We have obtained deep images in F555W and F814W 
of UGC~2302 using the WFPC2 on HST.  The images were reduced, calibrated, 
and aligned.  Photometry was performed on the stars common to both filters.
We find: (1) UGC~2302 clearly shows underlying spiral structure (see Figure 1, 
left); (2) UGC~2302 has a small nucleus or central star cluster (see Figure 1,
left); (3) The disk of UGC~2302 appears to be nearly transparent except in the
very central regions (see Figure 1, left); and (4) We have detected the tip 
of the Red Giant Branch (TGRB) at m$_{F555W} \sim 25$ and 
m$_{F555W}-$m$_{F814W} \sim 1.5$ (see Figure 1, right).  Also seen is the
blue main sequence at m$_{F555W} \sim 25$ and m$_{F555W}-$m$_{F814W} \sim 0$.

%\begin{enumerate}
%\item{UGC~2302 clearly shows underlying spiral structure (see Figure 1, left).}
%\item{UGC~2302 has a small nucleus or central star cluster (see Figure 1, left).}
%\item{The disk of UGC~2302 appears to be nearly transparent except in the very 
%central regions (see Figure 1, left).}
%\item{We have detected the tip of the Red Giant Branch (TGRB) at 
%m$_{F555W} \sim 25$ and m$_{F555W}-$m$_{F814W} \sim 1.5$ 
%(see Figure 1, right).  
%Also seen is the 
%blue main sequence at m$_{F555W} \sim 25$ and m$_{F555W}-$m$_{F814W} \sim 0$.}
%\end{enumerate}

\end{document}